\documentstyle[12pt,epsf]{article}
\def\o{\over}

\def\ra{\rightarrow}
\def\t{\tilde}

\def\be{\begin{equation}}
\def\ee{\end{equation}}
\def\bea{\begin{eqnarray}}
\def\eea{\end{eqnarray}}
\def\lsim{\mathrel{\lower2.5pt\vbox{\lineskip=0pt\baselineskip=0pt
           \hbox{$<$}\hbox{$\sim$}}}}
\def\gsim{\mathrel{\lower2.5pt\vbox{\lineskip=0pt\baselineskip=0pt
           \hbox{$>$}\hbox{$\sim$}}}}
\begin{document}
\setlength{\baselineskip}{8mm}
\begin{titlepage}
\begin{flushright}
\begin{tabular}{c c}
& {\normalsize  DPNU-96-05} \\
& {\normalsize February 1996}
\end{tabular}
\end{flushright}
\vspace{5mm}
\begin{center}
{\large \bf Higgs Masses and $CP$ Violation in SUSY Models} \\
\vspace{15mm} 
Naoyuki HABA\footnote{E-mail:\ haba@eken.phys.nagoya-u.ac.jp} \\ 

{\it 
Department of Physics, Nagoya University \\
           Nagoya, JAPAN 464-01 \\
}
\end{center}

\vspace{10mm}


\begin{abstract}

Though the mass of Higgs particle is the parameter 
determined by the experiment in the standard model (SM), 
SUSY models have rather predictive power for the 
lightest Higgs mass, and its upper bound in some SUSY models are 
close to the observable region in LEP2.   
The upper bound of the lightest Higgs mass is 
analysed systematically on the basis of the $CP$ violation in 
the minimal and the next 
minimal supersymmetric standard model (MSSM and NMSSM). 
In the explicit $CP$ violation case, the mass bound 
is large around $130 \sim 160$ GeV in both models. 
In the spontaneous $CP$ violation case induced by the 
radiative effects, 
the lightest Higgs mass upper bound is about $52$ GeV and 
sum of two light neutral Higgs should be around $O$(100 GeV) 
in the NMSSM in contrast to the one 
in the MSSM which implies about 6 GeV. 
This model gives the interesting predictions 
for the neutron electric dipole moment. 
\end{abstract}
\end{titlepage}

%
\section{Introduction}
The $CP$ physics is one of the most exciting 
topics in the recent particle physics. 
The origin of $CP$ violation is 
still in a mystery. 
In the standard model (SM), the 
origin of $CP$ phase exists in 
Kobayashi-Maskawa (KM) matrix\cite{KM}. 
In the SM, $CP$ is violated through the yukawa coupling 
and is conserved in the Higgs potential. 
The SM Lagrangian is not invariant under the 
$CP$ transformation and $CP$ is violated explicitly. 
However, if the Higgs sector is extended into the one with 
two or more doublets, 
we have richer $CP$ violation sources. 
In the multi-Higgs models, 
$CP$ is generally violated explicitly and/or spontaneously 
in the Higgs potential\cite{THDM}\cite{THDM2}. 
In the spontaneous $CP$ violation, 
vacuum expectation values (VEVs) have the non-trivial phases and 
the vacuum is not $CP$ invariant even if 
Lagrangian is $CP$ invariant. 
The simple model, in which $CP$ violation can occur 
explicitly and spontaneously, 
is the two Higgs doublet model (THDM). 
Since the THDM induces large flavor changing 
neutral current (FCNC) in general, 
one imposes some additional symmetries such as  
discrete symmetry\cite{DISCRITE} or 
approximate global family symmetry\cite{WOLFEN} on the model. 
\par
SUSY models automatically avoid large FCNC because 
one Higgs doublet ($H_1$) couples with down-sector 
and another ($H_2$) couples with up-sector. 
Additional parameters such as soft SUSY breaking parameters 
could be the origin of $CP$ violation and 
$CP$ would be violated both explicitly and spontaneously 
in the SUSY models. 
The Higgs bosons are the most important particles which 
the experimentalists and theorists wait for observing. 
In this paper, we study the Higgs masses in 
the minimal and the next minimal supersymmetric standard 
model\cite{NMSSM} 
(MSSM and NMSSM) 
with respect to the origin of $CP$ violation. 
It is found that the masses strongly depend on whether $CP$ is 
violated explicitly or spontaneously. 
And even in the explicit breaking case, the masses also depend on 
whether Higgs potential breaks $CP$ symmetry or not. 
\par
Let us give the brief review 
of the MSSM and the NMSSM. 
If we take into account of only top yukawa coupling for the 
yukawa sector, the superpotential of the MSSM and the NMSSM are 
\begin{equation}
\label{WMSSM}
W = h_tQH_2T^c + \mu H_1 H_2 , 
\end{equation}
and   
\begin{equation}
\label{WNMSSM}
W = h_tQH_2T^c + \lambda N H_1 H_2 - {k \over 3} N^3 , 
\end{equation}
respectively. 
Here $H_1$ and $H_2$ are Higgs doublet fields as 
\begin{equation}
H_1=\left(
\begin{array}{c}
H_1^{0} \\
H_1^- \\
\end{array}
\right) , \qquad
H_2=\left(
\begin{array}{c}
H_2^+ \\
H_2^0 \\
\end{array}
\right).
\end{equation}
with
\begin{equation}
  H_1H_2=H_1^0H_2^0-H_1^-H_2^+. 
\end{equation}
$Q$ is a third generation quark doublet superfield and 
$T^c$ is a right-handed top quark superfield. 
Top yukawa coupling constant is denoted by $h_t$. 
$N$ is a gauge singlet field. 
We neglect linear and quadratic terms of $N$ because 
of imposing $Z_3$ symmetry which might interpret 
weak scale baryogenesis\cite{BARYON}. 
If we include all these terms, the following 
discussion becomes quite 
different due to the additional parameters\cite{POM}. 
The parameters in the MSSM such as
\begin{equation}
\label{parameterMSSM}
 \mu, \;\; A_t,\;\; B,\;\; M_i \; (i = 1,2,3), 
\end{equation}
are complex in general. 
$A_t$ and $B$ terms are soft SUSY breaking parameters 
corresponding to 
top yukawa coupling and $\mu$ term, respectively. 
$M_i$s are gaugino mass parameters with the gauge group index $i$. 
By the $R$ transformation and Higgs field redefinition, 
two complex phases among four phases in Eq.(\ref{parameterMSSM}) 
can be rotated away. 
Then it is noticed that the $CP$ phases other than KM phase exist 
in the MSSM\cite{CPMSSM}. 
As for the NMSSM, adding parameters 
\begin{equation}
\label{parameterNMSSM}
 \lambda, \;\; k,\;\; A_{\lambda}, \;\; A_k, 
\end{equation}
are also generally complex, where  
$A_{\lambda}$ and $A_k$ are soft SUSY breaking parameters 
corresponding to $\lambda$ and $k$ in Eq.(\ref{WNMSSM}). 
The NMSSM has more $CP$ phases than the MSSM. 
\par
Section 2 is devoted to the explicit $CP$ violation through the 
yukawa sector. 
In section 3, we discuss the explicit $CP$ violation through the 
Higgs sector. 
In section 4, the spontaneous $CP$ violation and the neutron 
electric 
dipole moment (NEDM) are analysed. 
Section 5 gives summary and discussion. 
%
%
\section{The explicit $CP$ violation through the "yukawa" sector}
In general, explicit $CP$ violation 
occurs through the yukawa sector and/or the Higgs sector. 
In the former scenario, $CP$ violation should be 
induced by the yukawa couplings or scalar 
three point interactions, which we call "yukawa" sector 
$CP$ violation, analysed in this section. 
The latter scenario, where the Higgs sector breaks $CP$ 
symmetry explicitly, is discussed in section 3. 
\par
In this case, there is no mixing among scalar and pseudoscalar 
Higgs particles in the MSSM and NMSSM and 
the neutral Higgs mass matrix is  
\be
\label{matCPyukawa}
 M_{H^0}^2 = \bordermatrix{
   & {\rm Re}[H_1, H_2, (N)] & {\rm Im}[H_1, H_2, (N)]    \cr 
   & {\rm (scalar)}          &    0                 \cr
   &    0                    & {\rm (pseudoscalar)} \cr 
  } \ .
\ee
One of the pseudoscalars which is the mixing state of $H_1$ and 
$H_2$ is the Goldstone boson absorbed by $Z$ boson. 
There are two (three) neutral scalars and one (two) 
neutral pseudoscalar(s) in the MSSM (NMSSM) as physical particles.
\par
It is well known that the one loop 
corrections have non-negligible effects on 
Higgs masses in SUSY models\cite{massboundMSSM}. 
The one loop effective potential\cite{COLMAN} is 
\begin{equation}
\label{effect}
V_{1-{\rm loop}} = {1 \over 64 \pi^2} {\rm Str} \: M^4 ({\rm ln} \: 
                                                 {M^2 \over Q^2}),
\end{equation}
where $M$s are the field dependent mass matrices. 
If we consider only top and stop contributions and 
also neglect the stop left-right mixing, 
Eq.(\ref{effect}) is reduced to be 
\begin{equation}
\label{topeffect}
V_{\rm top} = {3 \over 64 \pi^2} 
    \left[ (h_t^2 |H_2|^2 + m_{\t t}^2)^2 {\rm ln} 
          {(h_t^2 |H_2|^2 + m_{\t t}^2) \over Q^2} - 
        h_t^4 |H_2|^4 {\rm ln} {h_t^2 |H_2|^2 \over Q^2} \right] \ ,
\end{equation}
where $m_{\t t}$ is the soft breaking stop mass. 
\par
By using Eq.(\ref{topeffect}), 
one can derive the upper bound of the lightest scalar masses 
both in the MSSM\cite{massboundMSSM} and 
in the NMSSM\cite{massboundNMSSM}\cite{Deer}\cite{ELLIS} as 
\be
\label{upMSSM}
 m_{h1} \leq M_Z^2 \; \cos^2 2 \beta + 
\Delta \; v^2 \sin^4 \beta \;  ,
\ee
and 
\be
\label{upNMSSM}
 m_{h1} \leq M_Z^2 \; \cos^2 2 \beta + 
                   {\lambda^2 \over 2} v^2 \; \sin^2 2 \beta + 
                             \Delta \; v^2 \sin^4 \beta \; ,
\ee
respectively. 
Here $\Delta$ is defined as 
\begin{equation}
\label{DELTA} 
\Delta \equiv {3 h_t^4 \over 4 \pi^2} \; {\rm ln}\: 
                                   {m_{\t t}{}^2 \over m_t^2} , 
\end{equation}
and VEVs of Higgs fields are 
\be
\label{VEV1}
\langle H_1 \rangle=v_1, \;\;\;\;\; \langle H_2 \rangle=v_2, 
\;\;\;\;\; 
\langle N \rangle=x.
\ee
Here $v_1, v_2$, and $x$ are real and positive parameters with 
$v=\sqrt{v_1^2+v_2^2}=174$ GeV. 
We also define $\tan \beta = v_2 / v_1$. 
The case of VEVs having non-vanishing relative phases is 
discussed later as the spontaneous $CP$ violation scenario. \\
It is worth noting that Eqs.(\ref{upMSSM}) and (\ref{upNMSSM}) 
do not change drastically by introducing more 
doublet- or singlet-Higgs fields\cite{Deer}. 
\par
Eq.(\ref{upMSSM}) shows that 
the MSSM light Higgs mass becomes too small 
in the region $\tan \beta \sim 1$. 
At $\tan \beta = 1$ first term is vanished and the 
loop effects play the essential role to lift up the Higgs mass. 
However the situation is quite different in the NMSSM. 
The second term in the Eq.(\ref{upNMSSM}) still works 
around $\tan \beta \sim 1$ and 
the scalar mass is not so small in 
contrast to the MSSM case. 
At large $\tan \beta$, 
the behaviors of scalar mass bound are almost the 
same in the MSSM and the NMSSM. 
It is also noticed that the larger stop mass becomes, 
the larger the light Higgs mass becomes 
from Eqs.(\ref{upMSSM}) and (\ref{upNMSSM}) . 
These behaviors are shown in Fig.1(a) and Fig.1(b). 
Through this paper we consider the case of 
$\tan \beta \geq 1$. \\
The lower bound of Higgs mass of the SM\cite{massboundSM} 
is also shown in the Fig.1(a) and Fig.1(b). 
This lower bound is obtained from the SM vacuum stability 
and written as 
\begin{equation}
\label{SMbound}
 m_{H_{\rm SM}} > 132.0 + 2.2 ( m_t - 170.0) - 
             {4.5 (\alpha_s - 0.117) \over 0.007} \;\; ({\rm GeV}).
\end{equation}
Here we use $\alpha_s = 0.129$, which is 
the strong coupling constant at $M_Z$ scale. 
%
\section{The explicit $CP$ violation through the Higgs sector}
In this section, we discuss the case that 
there is explicit $CP$ violation in the Higgs sector. 
Contrary to the previous situation Eq.({\ref{matCPyukawa}), 
the neutral Higgs mass matrix becomes 
\be
\label{matCPhiggs}
 M_{H^0}^2 = \bordermatrix{
   & {\rm Re}[H_1, H_2, (N)] & {\rm Im}[H_1, H_2, (N)]      \cr 
   & {\rm (scalar)}          &  \sim \sin \phi         \cr 
   & \sim \sin \phi          & {\rm (pseudoscalar)}       \cr 
  } \ .
\ee
Here $\phi$ is the phase that characterizes the $CP$ violation 
in the Higgs sector. 
\par 
In the MSSM, the tree level Higgs potential is automatically 
$CP$ invariant. 
$CP$ symmetry is violated by the radiative effects both 
explicitly and/or spontaneously. 
As for the spontaneous $CP$ violation case, 
we will see in the next section. 
In the explicit $CP$ violation case, 
the coefficients of $\lambda_{5,6,7}$ in Ref.\cite{THDM} 
derived by the radiative corrections are relatively small 
such as $\lambda_5 \simeq g^4/32 \pi^2 \sim 10^{-4}$ due to 
the loop suppression factor\cite{MAEKAWA}. 
The scalar-pseudoscalar mixing elements $S_1$-$A$ and $S_2$-$A$ 
of the neutral Higgs mass matrix are 
\begin{eqnarray}
\label{lam567}
 m^2_{S_1{\rm -}A} &=& \cos \beta \;{\rm Im}(B \mu) + 
        ( \sin^3 \beta - 2 \cos^2 \beta \sin \beta ) 
        \;{\rm Im}( \lambda_5 ) \; v^2      \nonumber            \\
   & & +( \cos \beta \sin^2 \beta - {3\over2}\cos^3 \beta ) \;
         {\rm Im}( \lambda_6 ) \; v^2 
  -{1\over2}\cos \beta \sin^2 \beta  \;
                   {\rm Im}( \lambda_7) \; v^2 \; , \nonumber  \\
 m^2_{S_2{\rm -}A} &=& - \sin \beta \;{\rm Im}(B \mu) + 
        ( 2 \cos \beta \sin^2 \beta - \cos^3 \beta ) 
        \;{\rm Im}( \lambda_5 ) \; v^2                          \\
   & & +{1\over2}\cos^2 \beta \sin \beta \;{\rm Im}( \lambda_6) \; v^2  
       +({3\over2}\sin^3 \beta - \cos^2 \beta \sin \beta ) \;
                {\rm Im}( \lambda_7 ) \; v^2  \;.      \nonumber           
\end{eqnarray}
Here we can always take $B \mu$ to be real 
by the Higgs field redefinition. 
Then the scalar-pseudoscalar mixing are so small 
that the situation becomes almost the same as the previous section. 
Then we go to the NMSSM following Ref.\cite{MATSUDA}. 
\par
The scalar-potential of the NMSSM including top, stop loop effects 
by Eq.(\ref{topeffect}) is 
\begin{equation}
V=V_{\rm no phase}+V_{\rm phase}\ , 
\end{equation}
where
\begin{eqnarray}
\label{phase}
 V_{\rm no phase} &=& |\lambda|^2 [|H_1 H_2|^2 + |N|^2 
( |H_1|^2 + |H_2|^2 )] +
                    |k|^2 |N|^4  \nonumber                   \\
   & & + \;{g_1^2 + g_2^2 \over 8}(|H_1|^2-|H_2|^2)^2 + 
        {g_2^2 \over 2} (|H_1|^2 |H_2|^2 - |H_1 H_2|^2)  
 \nonumber   \\
   & & + \; m_{H_1}^2 |H_1|^2 
       + m_{H_2}^2 |H_2|^2 + m_N^2 |N|^2                          
    \\ 
   & & + \; V_{\rm top} ,                                 
   \nonumber   \\
 V_{\rm phase}    &=& -(\lambda k^* H_1 H_2 N^{*2} + {\rm h.c.}) 
         -(\lambda A_{\lambda} H_1 H_2 N + {\rm h.c.}) -         
          ({k A_k \over 3} N^3 + {\rm h.c.})  \ .  \nonumber
\end{eqnarray}
The parameters $\lambda, k, A_\lambda$, and $A_k$ 
are all complex in general. 
$CP$ phase cannot be included in 
the potential corrected by the loop effect $V_{\rm top}$. 
So $CP$ phase appears from only $V_{\rm phase}$ 
in Eq.(\ref{phase}). 
We can remove two complex phases by 
the field redefinition of $N$ and $H_1 H_2$. 
So without loss of generality, we can take 
\begin{equation}
\lambda A_{\lambda} > 0, \;\;\;\;\; k A_k > 0. 
\end{equation}
Only one phase remains in $\lambda k^*$ denoted as 
\begin{equation}
 \lambda k^* \equiv \lambda k e^{i \phi}. 
\end{equation}
Here $\lambda$ and $k$ on the right hand side are 
real and positive numbers. 
In addition to this phase, there appear $CP$ phases from 
VEVs of $H_1, H_2$, and $N$ in general. 
But now, we neglect these phases for simplicity 
and use Eq.(\ref{VEV1}). 
By using the three stationary conditions 
\begin{equation}
\label{mini1}
 {\partial V \over \partial v_i} = 0 \; (i=1,2), \;\;\;\;\;\; 
 {\partial V \over \partial x} = 0, 
\end{equation}
we can eliminate three parameters $m_{H_1}^2, m_{H_2}^2$, 
and $m_N^2$. 
Higgs fields are expanded around their minimum point as 
\bea
\label{tenkai1}
 H_1^0 &=& v_1 + {1 \o \sqrt{2}} (S_1+i\sin\beta A) ,  
    \nonumber \\
 H_2^0 &=& v_2 + {1 \o \sqrt{2}} (S_2+i\cos\beta A) ,    
            \\
 N     &=& x + {1 \o \sqrt{2}} (X+iY).               
      \nonumber
\eea
Here $S_1, S_2$, and $X$ are scalars, 
and $A$ and $X$ are pseudoscalars. 
By using this notation, we get $5 \times 5$ 
neutral Higgs mass matrix as 
\begin{equation}
\label{mat5*5}
 M_{H^0}^2 = 
 \left(
  \begin{array}{cc}
   M^{S_1,S_2,X}_{S_1,S_2,X}    & M^{A,Y}_{S_1,S_2,X}       \\ 
                                &                           \\
   (M^{A,X}_{S_1,S_2,X})^T      & M^{A,Y}_{A,Y}       
  \end{array}
 \right) \;,
\end{equation}
where $M^{S_1,S_2,X}_{S_1,S_2,X}$, $M^{A,Y}_{S_1,S_2,X}$, and 
$M^{A,Y}_{A,Y}$ are $3 \times 3$, $3 \times 2$, and $2 \times 2$ 
submatrices, respectively. 
This matrix has the same form as Eq.(\ref{matCPhiggs}).
The matrix $M^{S_1,S_2,X}_{S_1,S_2,X}$ of 
the scalar part of $S_1, S_2$, and 
$X$ is
\begin{equation}
\label{mat3*3}
 M^{S_1,S_2,X}_{S_1,S_2,X} = 
 \left(
  \begin{array}{lll}
   \overline{g}^2 v^2 \cos^2 \beta  & 
        (\lambda^2 - \overline{g}^2/2) v^2 \sin 2 \beta &
                                2 \lambda^2 v x \cos \beta \\
   + \lambda x A_{\sigma_1} \tan \beta  & 
    -\lambda x A_{\sigma_1} &  
-\lambda v \sin \beta A_{\sigma_2} \\
                                     &  &       \\
   (\lambda^2 - \overline{g}^2/2) v^2 \sin 2 \beta & 
                  (\overline{g}^2+ \Delta) v^2 \sin^2 \beta &
                                2 \lambda^2 v x \sin \beta  \\
   - \lambda x A_{\sigma_1}            &
                  + \lambda x A_{\sigma_1}/ \tan \beta &
                            -\lambda v \cos \beta A_{\sigma_2} \\ 
                                     &  &       \\
   2 \lambda^2 v x \cos \beta         &
                  2 \lambda^2 v x \cos \beta        &
            {\lambda v^2 \over 2 x} A_{\lambda} \sin 2 \beta \\
         - \lambda v \sin \beta A_{\sigma_2} &
                  - \lambda v \cos \beta A_{\sigma_2} &
                        - A_k k x + 4 k^2 x^2                             
  \end{array}
 \right) \;,
\end{equation}
where we define $\overline{g}^2 \equiv (g^2+g'^2)/2$, 
$A_{\sigma_1} \equiv A_{\lambda} + k x \cos \phi$, and 
$A_{\sigma_2} \equiv A_{\lambda} + 2 k x \cos \phi$. 
The matrix $M^{A,Y}_{A,Y}$ of the pseudoscalar part of $A$ and 
$Y$ is
\begin{equation}
\label{mat2*2}
 M^{A,Y}_{A,Y} = 
 \left(
  \begin{array}{cc}
   2 \lambda x A_{\sigma_1}/ \sin 2 \beta    & 
 \lambda v A_{\sigma}'   \\ 
                                       &       \\
   \lambda v A_{\sigma}'               & 
   {\lambda v^2 \over 2 x} \sin 2 \beta A_{\lambda} + 3 A_k k x  \\
            &   + 2 \lambda k v^2 \sin 2 \beta \cos \phi     
  \end{array}
 \right) \;,
\end{equation}
where we define $A_{\sigma}' \equiv A_{\lambda} - 2 k x \cos \phi$. 
The matrix $M^{A,Y}_{S_1,S_2,X}$ of the 
scalar-pseudoscalar mixing part of $A$ and $Y$ is 
\begin{equation}
\label{mat2*3}
 M^{A,Y}_{S_1,S_2,X} = 
 \left(
  \begin{array}{cc}
   \lambda k x^2 \cos \beta \sin \phi 
                   &  -2 \lambda k v x \sin \beta \sin \phi \\
   \lambda k x^2 \sin \beta \sin \phi 
                   &  -2 \lambda k v x \cos \beta \sin \phi \\
   2 \lambda k v x \sin \phi         
                   &  - \lambda k v^2 \sin 2 \beta \sin \phi 
  \end{array}
 \right) \;.
\end{equation}
At $\phi = 0$, $M^{A,Y}_{S_1,S_2,X}$ vanishes and the 
Higgs mass matrix reduces to the type of Eq.(\ref{matCPyukawa}).
So the light Higgs mass becomes the same as the one 
in the "yukawa" CP violation case and 
$CP$ is conserved in the neutral Higgs sector. 
\par 
Now we consider the large $\tan \beta$ limit. 
$S_1$-$A$, $S_2$-$Y$, and $X$-$Y$ components in the 
$M_{S_1,S_2,X}^{A,Y}$ vanish at this limit. 
So it is enough to see the $S_1$-$Y$ and $S_2$-$X$-$A$ submatrices. 
These are 
\begin{equation}
\label{matS1Y}
 \left(
  \begin{array}{cc}
   \sim \lambda x A_{\sigma_1} \tan \beta & 
-2 \lambda k v x \sin \phi \\
   -2 \lambda k v x \sin \phi             & 3 A_k k x      
  \end{array}
 \right) \; ,
\end{equation}
and 
\begin{equation}
\label{matS2XA}
 \left(
  \begin{array}{ccc}
   (\overline{g^2} + \Delta) v^2  &  2 \lambda^2 v x  
                         & \lambda k x^2 \sin \phi   \\
   2 \lambda^2 v x                  &  -A_k k x + 4 k^2 x^2 
                            & 2 \lambda k v x \sin \phi \\
   \lambda k x^2 \sin \phi           &  2 \lambda k v x \sin \phi 
                  & 2 \lambda x A_{\sigma_1}/ \sin 2 \beta \\
  \end{array}
 \right) \;,
\end{equation}
respectively. 
In each matrix, only $S_1$-$S_1$  and $A$-$A$ components 
are dominant and the scalar-pseudoscalar mixing is very small. 
$CP$ violation in the neutral Higgs sector vanishes 
at the large $\tan \beta$ limit . 
In this case, 
the light Higgs mass is the same as explicit $CP$ violation 
in the "yukawa" sector. 
\par 
The scalar-pseudoscalar mixing depends on the value 
of $\tan \beta$. 
In the region of $\tan \beta \sim 1$, 
this mixing becomes large. 
Then the light scalar mass become smaller than 
the one without mixing. 
The $\tan \beta$ dependence of the light 
scalar mass is shown in Fig.2, in which we take 
from Ref.\cite{MATSUDA}, as 
\begin{eqnarray}
 & &  k = 0.1,\;\;\;\;\; \lambda = 0.2, \;\;\;\;\; 
 m_{\t t} = 3 \;{\rm TeV}, \nonumber \\
 & &  A_k = A_{\lambda} = v, \;\;\;\;\; x = 10 \; v.
\end{eqnarray}
The Higgs particle gets smaller mass as the phase $\phi$ 
becomes larger. 
In Fig.2, the following experimental constraints of Higgs 
search are 
considered.; 
\begin{enumerate}
\item The lightest and 
the second lightest Higgs bosons denoted by 
$h_1$ and $h_2$ have not been observed 
in the decay of $Z$\cite{PDG}, 
so that $Z \ra h_1+h_2$ should be forbidden 
kinematically. Then the condition
$$
m_{h_1}+m_{h_2} > m_Z
$$
is derived. 
\item The lightest boson $h_1$ has 
not been observed by the decay $Z \ra
h_1+Z^* 
\ra h_1 +l^+l^-$
\cite{ALEPH}.
The lower mass limit is 
$$
m_{h_1} > (65{\rm GeV})(\alpha_1\cos\beta+\alpha_2\sin\beta)^2,
$$
where $\alpha_1$ ($\alpha_2$) is the ratio of the $S_1$ ($S_2$) 
component of $h_1$. 
\item The "pseudoscalar" boson should be larger than 22 GeV 
in the case of $\tan \beta > 1$\cite{PDG}. 
\item In the MSSM, the lower limit of two Higgs scalars 
should be larger 
than 44 GeV in the case of $\tan \beta > 1$\cite{PDG}. 
\end{enumerate}
%
%
\section{The spontaneous $CP$ violation}
In this section, we discuss the spontaneous $CP$ violation in the 
MSSM and the NMSSM. 
It occurs by the phase difference of VEVs of Higgs fields. 
As for the MSSM, the tree level Higgs potential 
is always $CP$ invariant. 
However, as seeing in the previous section, 
if the radiative corrections are included, 
there is the possibility of the spontaneous 
$CP$ violation\cite{MAEKAWA}.
In this case, the light "pseudoscalar" appears with 
about $6$ GeV mass and this is contradict with experiment\cite{PDG}. 
In the NMSSM, 
if we consider only cubic couplings, the tree level 
potential cannot 
have $CP$ violating vacuum\cite{ROMAO}. 
However the one loop corrections could trigger spontaneous $CP$ 
violation\cite{BABU}. 
This scenario also demands the relatively light "pseudoscalar" 
compared to the no-$CP$ violation scenarios. 
The appearance of light particles in both the MSSM and the NMSSM 
is the general results by the 
Georgi-Pais theorem\cite{GP}. 
Since we study the possibility of the spontaneous $CP$ violation, 
all parameters except for VEVs of $H_1, H_2$, and $N$ are 
assumed to be real. 
We obtain the Higgs mass around 50 GeV, 
which is compatible with the present experimental constraints, 
because there are adjustable parameters in the NMSSM. 
\par 
Recently, Babu and Barr pointed out the possibility of 
the spontaneous $CP$ violation in the NMSSM 
by using Eq.(\ref{topeffect})\cite{BABU}. 
The results of their analysis are summarized as 
\begin{equation}
\label{BabuBarr}
 m_{h_1}^2 \leq C M_Z^2 , \;\;\;\;\; 
 m_{h_1}^2 + m_{h_2}^2 \leq (C + \cos^2 \beta) M_Z^2 \; , 
\end{equation}
where 
$C \equiv [4 A_{\lambda} 
(3 A_{\lambda}-A_k)/ A_k (4 A_{\lambda}-A_k)] 
(\lambda / \overline{g}^2)$. 
The positivity of the mass eigen-values and the experimental 
constraints limit the parameter to 
\begin{equation}
  1/3 \leq A_{\lambda}/A_k \leq 2.7 .
\end{equation}
So the upper bounds of Higgs masses are 
\begin{equation}
 m_{h_1} \leq 50 \;{\rm GeV}, 
\;\;\;\;\; m_{h_1}+m_{h_2} \leq 100 \;{\rm GeV}. 
\end{equation}
The predictive charged Higgs mass $m_{H^{\pm}} \leq 100$ GeV 
is not affected by 
the radiative corrections. 
As long as one uses Eq.(\ref{topeffect}), 
$CP$ phase does not appear in Eq.(\ref{effect}). 
However, if stop left-right mixing terms are included, 
which are neglected in Eq.(\ref{topeffect}), 
there also exists the $CP$ phase in the one loop level 
effective potential of Eq.(\ref{effect}). 
There is the possibility that the $CP$ violating 
effects which appear in the one loop level 
might have the large effects on the Higgs masses. 
As for this, numerical analysis has been done in Ref.\cite{HABA}, 
where bottom and sbottom contributions are 
also included. 
We use the squark 
mass squared matrix $M^2$ in Eq.(\ref{effect}) as 
\be
\label{emass}
M_{{\t t},{\t b}}^2 =\bordermatrix{
& \tilde{t}_L &{\tilde{t}}^c_R &\tilde{b}_L &{\tilde{b}}^c_R \cr 
& m_{11}^2 & m_{12}^2 & m_{13}^2 & m_{14}^2 \cr 
& m_{12}^{*2} & m_{22}^2 & m_{23}^2 & m_{24}^2 \cr 
& m_{13}^{*2} & m_{23}^{*2} & m_{33}^2 & m_{34}^2 \cr 
& m_{14}^{*2} & m_{24}^{*2} & m_{34}^{*2} & m_{44}^2 \cr } \ ,
\ee
where
\bea
m_{11}^2&=&m_Q^2 + h_t^2|H_2^0|^2+ 
h_b^2|H_1^-|^2-{g_1^2 \o 12}(|H_1^0|^2 
+|H_1^-|^2- |H_2^0|^2-|H_2^+|^2) \nonumber \\ 
&+&{g_2^2 \o 4}(|H_1^0|^2-|H_1^-|^2-|H_2^0|^2+|H_2^+|^2), 
\nonumber \\ 
m_{12}^2&=&h_t(A_tH_2^{0*}+\lambda NH_1^0), \nonumber \\ 
m_{13}^2&=&-h_t^2H_2^{0*}H_2^+-h_b^2H_1^{-*}H_1^0+
{g_2^2 \o 2}(H_2^+H_2^{0*}+H_1^{-*}H_1^0), \nonumber\\ 
m_{14}^2&=&-h_b(\lambda N H_2^{+}-A_bH_1^{-*}), \nonumber\\ 
m_{22}^2&=&m_T^2+h_t^2(|H_2^0|^2+|H_2^+|^2)+
{g_1^2 \o 3}(|H_1^0|^2 +|H_1^-|^2- |H_2^0|^2-|H_2^+|^2), 
\nonumber\\ 
m_{23}^2&=&h_t(\lambda N^*H_1^{-*}-A_tH_2^+), \nonumber\\ 
m_{24}^2&=&h_th_b(H_2^{0}H_1^{-*}+H_2^{+}H_1^{0*}), 
\nonumber\\ 
m_{33}^2&=&m_Q^2+h_b^2|H_1^0|^2+ 
h_t^2|H_2^+|^2-{g_1^2 \o 12}
(|H_1^0|^2 +|H_1^-|^2- |H_2^0|^2-|H_2^+|^2), \nonumber \\ 
&+&{g_2^2 \o 4}\left(-|H_1^0|^2
+|H_1^-|^2+|H_2^0|^2-|H_2^+|^2\right), \nonumber \\ 
m_{34}^2&=&-h_b(A_bH_1^{0*}+\lambda NH_2^0), \nonumber \\ 
m_{44}^2&=&m_B^2+h_b^2(|H_1^0|^2+|H_1^-|^2)-
{g_1^2 \o 6}(|H_1^0|^2 +|H_1^-|^2- |H_2^0|^2-|H_2^+|^2) \ . 
\nonumber
\eea
Here the mass parameters $m_Q,m_T$, and $m_B$ 
are the soft supersymmetry breaking 
squark masses, and the parameters 
$A_t$ and $A_b$ are the coefficients of 
the soft supersymmetry breaking 
terms as 
\be
V_{\rm soft}=
A_th_t{\tilde t}_L{\tilde t}^*_RH_2^0-A_bh_b{\tilde b}_L
{\tilde b}^*_RH_1^0+{\rm h.c.}+\cdots. 
\ee
By using Eq.(\ref{emass}), 
one loop contribution of the charged Higgs mass is obtained. 
And we can also analyze the large $\tan \beta$ region, in which 
$h_b \sim h_t$. 
The brief review is shown as follows. 
\par
In the spontaneous $CP$ violation case, 
there exists the phase difference of VEVs of Higgs fields in 
contrary to the Eq.(\ref{VEV1}). 
VEVs of Higgs fields are defined as 
\be
\label{VEV2}
 \langle H_1 \rangle=v_1 e^{i\varphi_1}, \;\;\;\;
 \langle H_2 \rangle=v_2 e^{i\varphi_2}, \;\;\;\; 
 \langle N \rangle=xe^{i\varphi_3}.
\ee
We can always eliminate one phase by 
the field redefinition. 
So physical phases are two which are assigned as 
\be
 \theta \equiv \varphi_1 + \varphi_2 + \varphi_3, \;\;\;\;\;
 \delta \equiv 3 \varphi_3.
\ee
The minimization conditions Eq.(\ref{mini1}) are modified to 
\begin{equation}
\label{mini2}
 {\partial V \over \partial v_i} = 0 \; (i=1,2), \;\;\;\; 
 {\partial V \over \partial x} = 0, \;\;\;\;
 {\partial V \over \partial \delta} = 0, \;\;\;\;
 {\partial V \over \partial \theta} = 0. 
\end{equation}
The soft breaking masses $m_{H_1}^2, m_{H_2}^2$, and $m_N^2$ 
are eliminated by the stationary conditions of $v_i$ and $x$. 
And $\delta$ and $k$ are eliminated by 
the stationary conditions of $\delta$ and $\theta$, respectively. 
\par
In addition to the experimental constraints $1 \sim 4$ 
in the previous section, we also impose the 
theoretical constraints\cite{ELLIS} 
\begin{equation}
|k| \leq 0.87, \;\;\;\; |k| \leq 0.63 .
\end{equation}
They are derived by the assumption that 
perturbation 
of $\lambda$ and $k$ should remain valid up to the GUT scale. 
Considering all these constraints, 
we obtain Higgs masses as the curvatures of the potential 
at the minimum point. 
The input parameters are 
\begin{eqnarray}
\label{parametersets}
 & &  \lambda = 0.24,\;\;\;\;  
            m_{{\t t_L},{\t b_L}} = 3 \; {\rm TeV}, \nonumber \\
 & &  m_{\t t_R} = 0.95 \; m_{\t t_L},\;\;\;\; 
               m_{\t b_R} = 0.98 \; m_{\t t_L} \; , \nonumber \\
 & &  A_t = 1 \; {\rm TeV}, 
\;\;\;\; A_b = 1.1 \; A_t \; ,       \nonumber \\
 & &  A_k = 20 
\; v,\;\;\;\; A_{\lambda} = 11 \; v,\;\;\;\; x = 20 \; v \; ,
\end{eqnarray}
with the assumption of GUT scale universality\cite{RG}. 
Here, $m_{{\t t_L},{\t b_L}}$, $m_{\t t_R}$, 
and $m_{\t b_R}$ are soft breaking masses. 
\par
In Ref.\cite{HABA}, we obtained the numerical results 
\begin{equation}
\label{habaresult}
 m_{h_1} \leq 52 \;{\rm GeV}, 
\;\;\;\;\; m_{h_1}+m_{h_2} \leq O(100 
                                         \;{\rm GeV}) , 
\end{equation}
in compatible with the experimental constraints $1 \sim 4$ 
given in the last section. 
These results are consistent 
with Ref.\cite{BABU}. 
So we can say that the neutral Higgs masses are not largely 
influenced by the $CP$ phase in the one loop potential. 
However, as for the charged Higgs mass, we obtain 
relatively large mass about $285$ GeV 
by the full one loop corrections including stop and sbottom 
mass matrix in Eq({\ref{emass}). 
The charged Higgs has too large mass to be 
observed at LEP2 and this
mass is enough large to be consistent with 
$b \rightarrow s\gamma$ experiment\cite{CLEO}.
\\
We have found that the solution only exists around 
the region in which $\tan \beta \simeq 1$, squark soft 
breaking masses 
are about 3 TeV, and $A_t$ and $A_b$ are about 1 TeV. 
\par
At present the neutron electric dipole moment (NEDM) 
gives the important clue to check the various 
$CP$ violation models beyond the SM. 
In the followings, we estimate the NEDM by using 
the parameters obtained in the spontaneous $CP$ 
violation scenario. 
The chargino and the gluino contributions to the NEDM 
are shown in Fig.3. 
The non-vanishing phases appear from 
the chargino and the squark mass matrices even if 
all initial parameters are set to be real. 
We assume that the gaugino masses satisfy 
the GUT relations as 
\begin{equation}
 {M_3 \over g_s^2} = {M_2 \over g^2} = {3 \over 5}{M' \over g'^2}, 
\end{equation}
where $M_3$, $M_2$, and $M'$ are soft breaking masses 
associated with the $SU(3)_C$, $SU(2)_L$, and $U(1)_Y$ subgroups, 
respectively. 
In the following calculations, 
we assume that $M_2$ is 1 TeV, 
sup and sdown masses are 3 TeV, and the flavor mixing is neglected 
for simplicity. 
By these assumptions, the chargino contribution (Fig.3 (a)) is 
larger than the gluino one (Fig.3 (b))\cite{OHSIMO}. 
The $CP$ phase $\theta$ appears 
from diagonalization of the chargino and 
the squark mass matrices, and 
we can not rotate away this phase by the field redefinition. 
{}For example, 
the EDM of down quark from the chargino contribution is 
\begin{eqnarray}
 d_d/e &=& {\alpha_{em} \over 4 \pi \sin^2 \theta_W} 
 \sin \theta {\lambda x M_2 \tan \beta  \over (m_{\omega_1}^2 
 - m_{\omega_2}^2)}
 {m_d \over m_{\t d}^2 }     \nonumber \\
 & & \times \sum_{i=1}^2 (-1)^i \left( -{1 \over 3} 
 I \left[ {m_{\omega_i^2} \over m_{\t d}^2} \right] + 
 J \left[ {m_{\omega_i^2} \over m_{\t d}^2} \right] \right) , 
\end{eqnarray}
where, 
\begin{eqnarray}
 & & I[r] \equiv {1 \over 2(1-r)^2} 
            \left( 1+r+{2r \over 1-r} {\rm ln} r \right) ,  \\
 & & J[r] \equiv {1 \over 2(1-r)^2} 
      \left( 3-r+{2r \over 1-r} {\rm ln} r \right) ,  \nonumber 
\end{eqnarray}
and $\omega_i$s are the chargino mass eigen-states. 
The gluino contribution is also estimated by the same way. 
The chargino and the gluino contributions are shown in Fig.4 by 
using the non-relativistic relation $d_n = (4 d_d - d_u)/3$. 
Spontaneous CP violation in the NMSSM predicts 
$d_n \simeq O(10^{-26})$ e$\cdot$cm which is small at one order 
compared 
to the present experimental upper limit of the NEDM\cite{PDG}. 
%
%
\section{Summary and Discussion}
We have studied the Higgs masses in the context with the 
$CP$ violation structures in the MSSM and the NMSSM. 
In the explicit $CP$ violation, 
the lightest Higgs mass can be large 
about $130 \sim 150$ GeV in both the MSSM and the NMSSM. 
This results do not change drastically whether 
$CP$ violation exists Higgs sector or not. 
We have shown that the problem of 
the spontaneous $CP$ violation scenario in the MSSM, 
which requires the light Higgs mass around 
a few GeV, is solved by extending the MSSM into the NMSSM 
with the singlet superfield. 
We also predict the NEDM which is 
smaller in one order than the present experimental upper limit. 
In this scenario, 
the lightest Higgs mass is $m_{h_1} \leq 52$ GeV and 
sum of two light Higgs masses should be around 100 GeV. 
It is expected that LEP2 experiment will give the answer for 
the possibility of spontaneous $CP$ violation 
in the Higgs sector.

\vskip 1 cm
\noindent
{\bf Acknowledgements}\par
I would like to thank M. Matsuda and M. Tanimoto for 
useful discussions and careful reading of manuscripts. 
I am also grateful to Y. Okada for useful comments and 
N. Oshimo for the discussions of NEDM. 


%
%
\newpage

\begin{center}
{\bf Figure Captions} \\
\end{center}
{\bf Fig.1}\qquad
The upper/lower bounds of Higgs masses versus $\tan \beta$. 
The parameter $\lambda$ = 0.4 is fixed. 
Solid line: 
the upper bound of the lightest scalar mass in the NMSSM;  
long-dashed-dotted line: the upper bound of the lightest 
scalar mass in the MSSM;  
dashed line: the lower bound of Higgs mass 
in the SM; \\
{}Fig.1(a):  $m_{\t t} = 1$ TeV;  \\
{}Fig.1(b):  $m_{\t t} = 3$ TeV.  \\
{\bf Fig.2} \qquad
The NMSSM lightest Higgs mass versus $\tan \beta$ with 
the explicit $CP$ violation in the Higgs potential. 
Solid line: $\phi= \pi/2$;  
dashed line: $\phi= \pi/4$;  
long-dashed-dotted line: $\phi= 0$. \\
{\bf Fig.3} \qquad
The diagram which contribute to the NEDM. \\
(a): The chargino contribution;  
(b): the gluino contribution. \\
{\bf Fig.4(a)} \qquad
The dependence 
of the NEDM on the phase $\theta$ at $\tan \beta =1$. 
The region of $\theta$ is where the spontaneous 
$CP$ violation in the NMSSM is available\cite{HABA}. 
Solid line: the chargino contribution;  
long-dashed-dotted line: the gluino contribution;  
dashed line: the experimental upper limit 
$11 \times 10^{-26}$ e$\cdot$cm. \\
{\bf Fig.4(b)} \qquad
The $x= \langle N \rangle$ dependence of the NEDM at $\theta = 1.7$. 
The region of $x$ is where the 
spontaneous $CP$ violation in the NMSSM is available. 
Solid line: the chargino contribution;  
long-dashed-dotted line: the gluino contribution;  
dashed line: the experimental upper limit. 
\vspace {5mm}
%
%
%
\newpage
{\bf Table} \qquad
The light Higgs masses for models (MSSM or NMSSM) with 
origins of $CP$ violation. 
\begin{center}

\begin{tabular}{|c|c|ccl|}  
\hline 
 Model & $CP$ violation &  &    &  
Mass bound                            \\
\hline 
\hline 
     &           &  
&           &                                        \\
     &           & 
"yukawa" & $m_{h_1}$ &  $\leq M_Z^2 \cos^2 2 \beta + 
                           \Delta v^2 \: \sin^4 \beta$   \\
     & explicit  &  &           &  
                $\leq 130 \sim 160$ 
GeV (small at $tan \beta \sim 1)$    \\
     &           &------------
&------ & --------------------------------------------------   \\
MSSM &          & Higgs &      
&  Almost same as "yukawa" case          \\
     & -------------- 
&------------&------ &
 --------------------------------------------------  \\
     &           &  &           & 
                                       \\
     &  spontaneous & 
& $m_{h_1}$ & $\leq 6$ GeV (excluded from experiment) \\
     &           &  
&           &                                        \\
\hline
     &         
  &  &           &                                        \\
     &       
    & "yukawa" & $m_{h_1}$ &  
$\leq M_Z^2 \cos^2 2 \beta +
{1 \over 2} \lambda^2 v^2 \sin^2 2 \beta + 
                            \Delta v^2 \: \sin^4 \beta$  \\
     &  explicit &        
  &   &  $\leq 130 \sim 160$ GeV               \\
     &           &------------
&------ & --------------------------------------------------   \\
NMSSM &          & Higgs    
& $m_{h_1}$ & $\leq O(150$ GeV)              \\
     & -------------- 
&------------&------ 
& --------------------------------------------------  \\
     &           &  &    
       &                                        \\
     & spontaneous  &  
     & $m_{h_1}$ & $\leq 52$ GeV                  \\
     &           &  
     & $m_{h_1}+m_{h_2}$ & $\leq O(100$ GeV)        \\
     &           &  &  
         &                                        \\
\hline
\end{tabular}

\end{center}


\newpage
\begin{center}
\epsfysize=10cm
\hfil\epsfbox{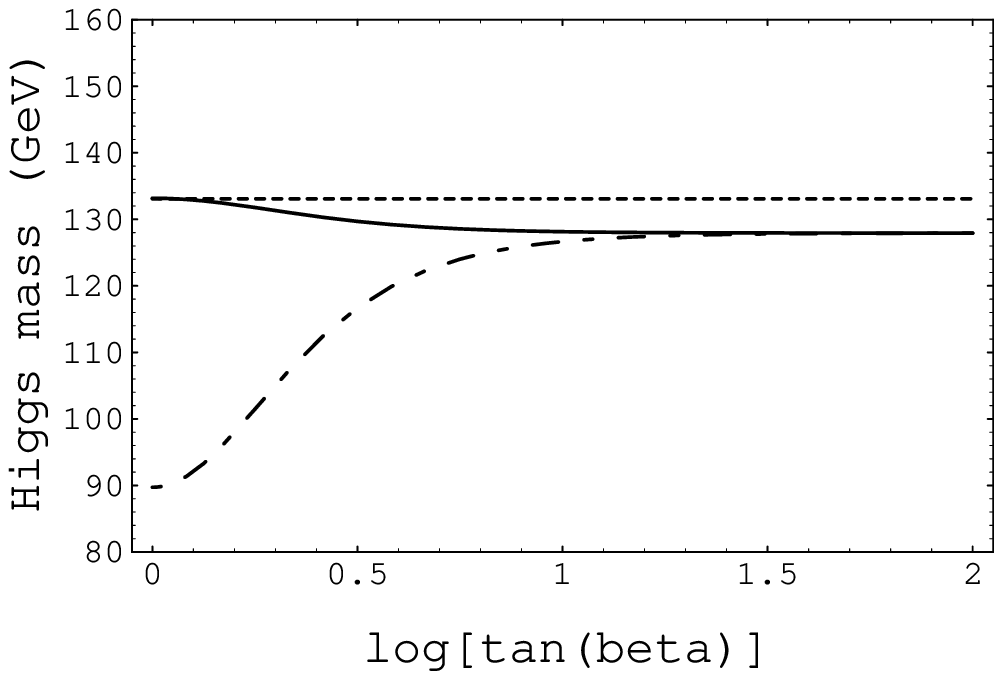}\hfill
\end{center}
\vspace{-3cm}
\begin{center}
{}Fig.1 (a) \vspace{1cm}\\
\end{center}

\begin{center}
\epsfysize=10cm
\hfil\epsfbox{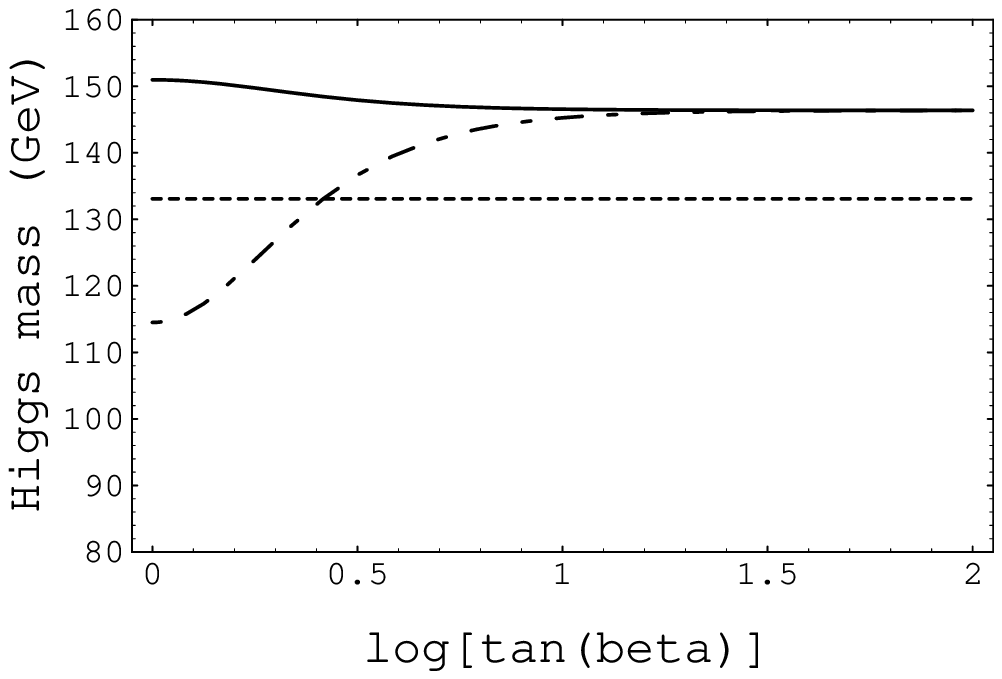}\hfill
\end{center}
\vspace{-3cm}
\begin{center}
{}Fig.1 (b)
\end{center}

\newpage
\begin{center}
\epsfysize=10cm
\hfil\epsfbox{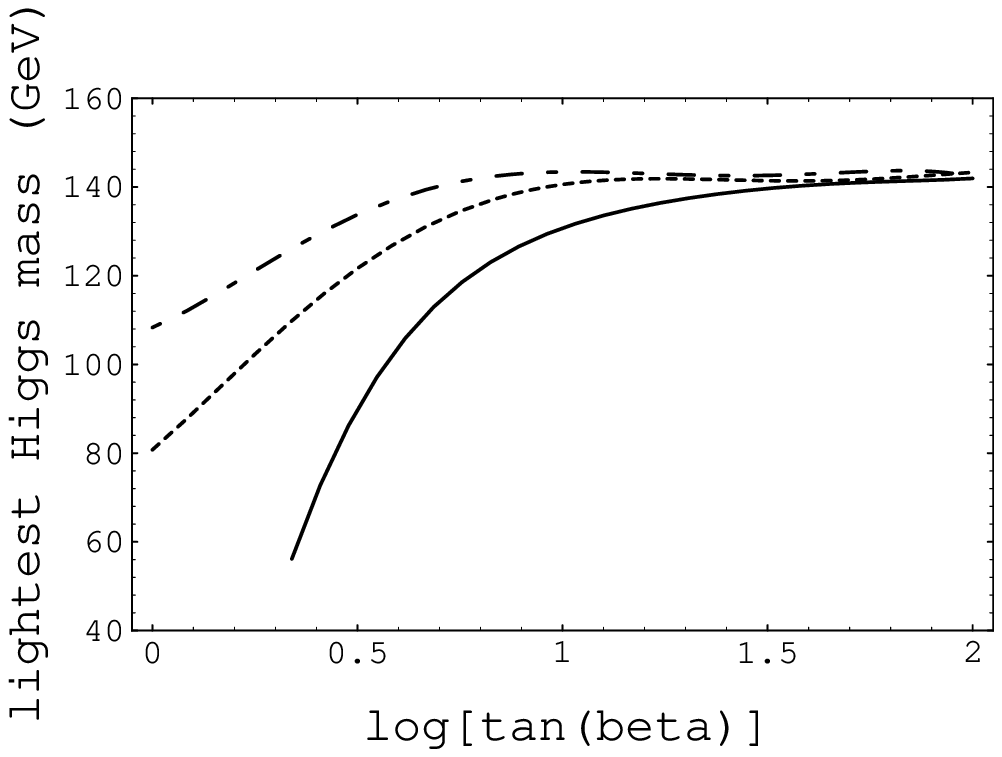}\hfill
\end{center}
\vspace{-2cm}
\begin{center}
{}Fig.2
\end{center}

\newpage
\vspace*{5cm}
\begin{center}
\epsfysize=10cm
\hfil\epsfbox{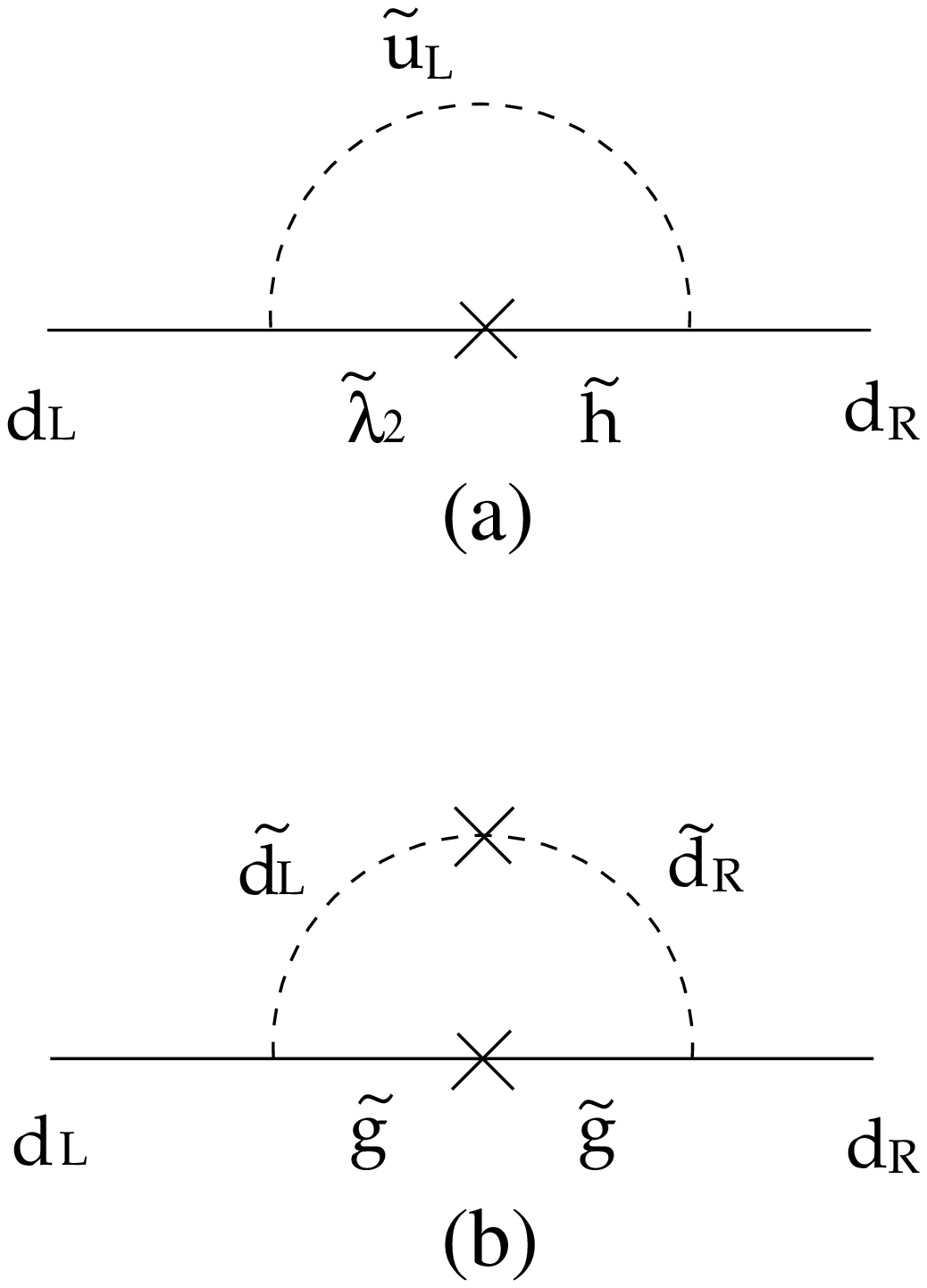}\hfill
\end{center}
\vspace{-3cm}
\begin{center}
{}Fig.3
\end{center}

\newpage
\begin{center}
\epsfysize=10cm
\hfil\epsfbox{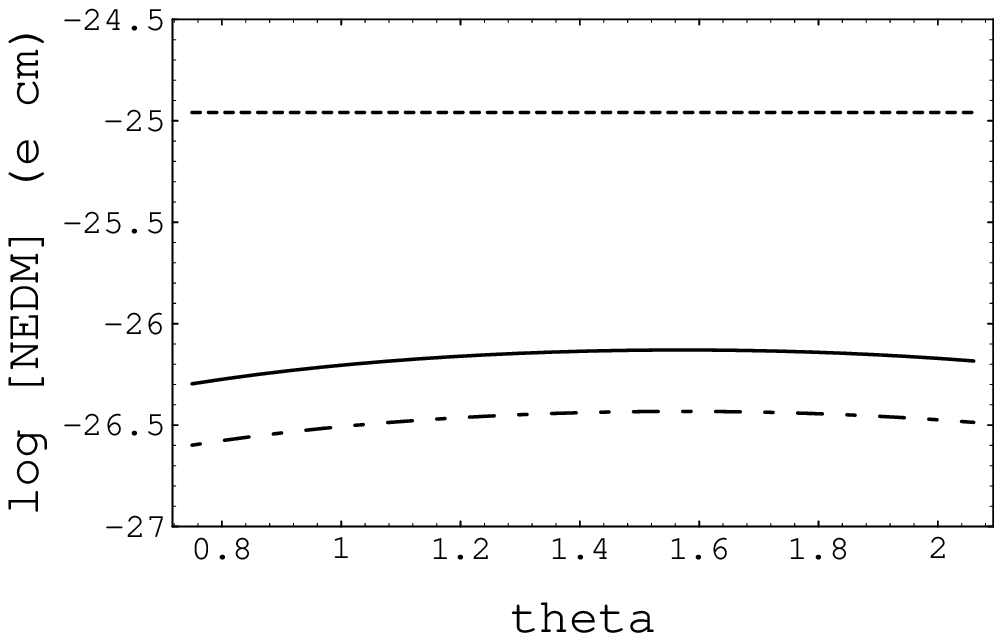}\hfill
\end{center}
\vspace{-3cm}
\begin{center}
{}Fig.4 (a) \vspace{1cm}\\
\end{center}

\begin{center}
\epsfysize=10cm
\hfil\epsfbox{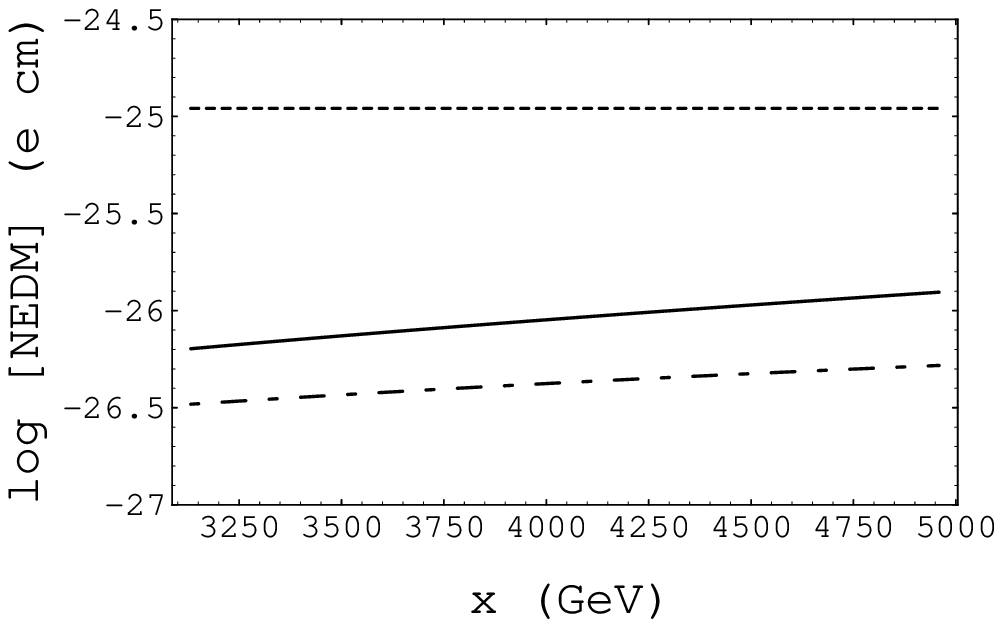}\hfill
\end{center}
\vspace{-3cm}
\begin{center}
{}Fig.4 (b)
\end{center}

\end{document}